\documentclass[pra, amsmath,amssymb,  superscriptaddress, reprint]{revtex4-1}%

\usepackage{dcolumn}
\usepackage{bm}
\usepackage{amsmath}
\usepackage{amsfonts}
\usepackage{amssymb}%
\usepackage{latexsym}
\usepackage{braket}
\usepackage{comment}

\usepackage{graphicx}
\usepackage{color}

\begin{document}

\newcommand{\change}[2]{{\color{blue} #1}{\color{red} #2}}

\preprint{{\large Preprint}} 

\title{Implementation of infinite-range exterior complex scaling to the time-dependent complete-active-space self-consistent-field method}

\author{Yuki Orimo}
\email{ykormhk@atto.t.u-tokyo.ac.jp}
\affiliation{Department of Nuclear Engineering and Management, School of Engineering, The University of Tokyo, 7-3-1 Hongo, Bunkyo-ku, Tokyo 113-8656, Japan}

\author{Takeshi Sato}
\affiliation{
Photon Science Center, School of Engineering, 
The University of Tokyo, 7-3-1 Hongo, Bunkyo-ku, Tokyo 113-8656, Japan}
\affiliation{Department of Nuclear Engineering and Management, School of Engineering, The University of Tokyo, 7-3-1 Hongo, Bunkyo-ku, Tokyo 113-8656, Japan}

\author{Armin Scrinzi}
\affiliation{Ludwig Maximilians Universit\"{a}t, Theresienstrasse 37, 80333 Munich, Germany}

\author{Kenichi L. Ishikawa}
\affiliation{Department of Nuclear Engineering and Management, School of Engineering, The University of Tokyo, 7-3-1 Hongo, Bunkyo-ku, Tokyo 113-8656, Japan}
\affiliation{
Photon Science Center, School of Engineering, 
The University of Tokyo, 7-3-1 Hongo, Bunkyo-ku, Tokyo 113-8656, Japan}


\begin{abstract}
We present a numerical implementation of the infinite-range exterior complex scaling (irECS) [Phys. Rev. A {\bf 81}, 053845 (2010)] as an efficient absorbing boundary to the time-dependent complete-active-space self-consistent field (TD-CASSCF) method [Phys. Rev. A {\bf 94}, 023405 (2016)] for multielectron atoms subject to an intense laser pulse. 
We introduce Gauss-Laguerre-Radau quadrature points to construct discrete variable representation basis functions in the last radial finite element extending to infinity.
This implementation is applied to strong-field ionization and high-harmonic generation in He, Be, and Ne atoms. It efficiently prevents unphysical reflection of photoelectron wave packets at the simulation boundary, enabling accurate simulations with substantially reduced computational cost, even under significant ($\approx 50\%$) double ionization. For the case of a simulation of high-harmonic generation from Ne, for example, 80\% cost reduction is achieved, compared to a mask-function absorption boundary.
\end{abstract}

\maketitle

\section{introduction}
Recent developments in ultrashort intense laser pulse techniques have opened new research fields including strong-field phenomena (e.g., tunneling ionization, high-harmonic generation (HHG), nonsequential double ionization) and ultrafast electronic dynamics (e.g., charge migration, photoemission delay).
Although these phenomena are rigorously described by the time-dependent Schr\"odinger equation (TDSE), solving it for multielectron systems poses a major challenge.  
To investigate many-electron dynamics in intense laser fields, time-dependent multiconfiguration self-consistent field (TD-MCSCF) methods have been developed \cite{Ishikawa_2015}, 
where the total wave function is given in the configuration interaction (CI) expansion,
\begin{gather}
	\Psi (\vec{x}_1, \vec{x}_2, \cdots, \vec{x}_N, t) = \sum_{I} C_I(t) \Phi_I(\vec{x}_1, \vec{x}_2, \cdots, \vec{x}_N, t).
\end{gather}
and $\vec{x}$ is a set of a spin coordinate $\sigma$ and spatial coordinate $\vec{r}$.
The electronic configuration $\Phi_I(\vec{x}_1, \vec{x}_2, \cdots, \vec{x}_N, t)$ is a Slater determinant composed of spin orbital functions $\{\psi_p(\vec{r},t) \times s(\sigma)\}$, where $\{ \psi_p(\vec{r},t) \}$ and $\{ s(\sigma) \}$ denote spatial orbitals and spin functions, respectively. 
Both the CI coefficients $\{C_I\}$ and orbitals are varied in time.
The multiconfiguration time-dependent Hartree-Fock (MCTDHF) method \cite{Zanghellini_2003,Kato_2004,Caillat_2005} considers all the possible configurations for a given number of orbital functions.
However, its computational cost factorially increases with the number of electrons.
To overcome this difficulty, we have recently developed and successfully implemented the time-dependent complete-active-space self-consitent-field (TD-CASSCF) method~\cite{Sato_2013}, which classifies the spatial orbitals into doubly occupied and time-independent \textit{frozen core (FC)}, doubly occupied and time-dependent \textit{dynamical core (DC)}, and fully correlated \textit{active} orbitals.
The number of configurations and the computational cost are significantly reduced without degrading accuracy.
We have further proposed a more approximate and thus computationally even less demanding
time-dependent occupation-restricted multiple active-space (TD-ORMAS) method \cite{Sato_2015}.

One of the key issues in numerical implementations of the TD-MCSCF methods is an absorbing boundary that absorbs the photoelectron wave packet when it reaches the end of the spatial grid and suppresses unphysical reflections.
An efficient absorbing boundary plays a significantly important role to achieve large scale simulations, for example, in simulations with three-dimensional Cartesian coordinates for general molecules \cite{Sawada_2016}, where the computational cost cubically scales with the linear dimension of the simulation box. 

\begin{figure}[t]
	\includegraphics[scale=0.5, clip, bb=0 0 491 283]{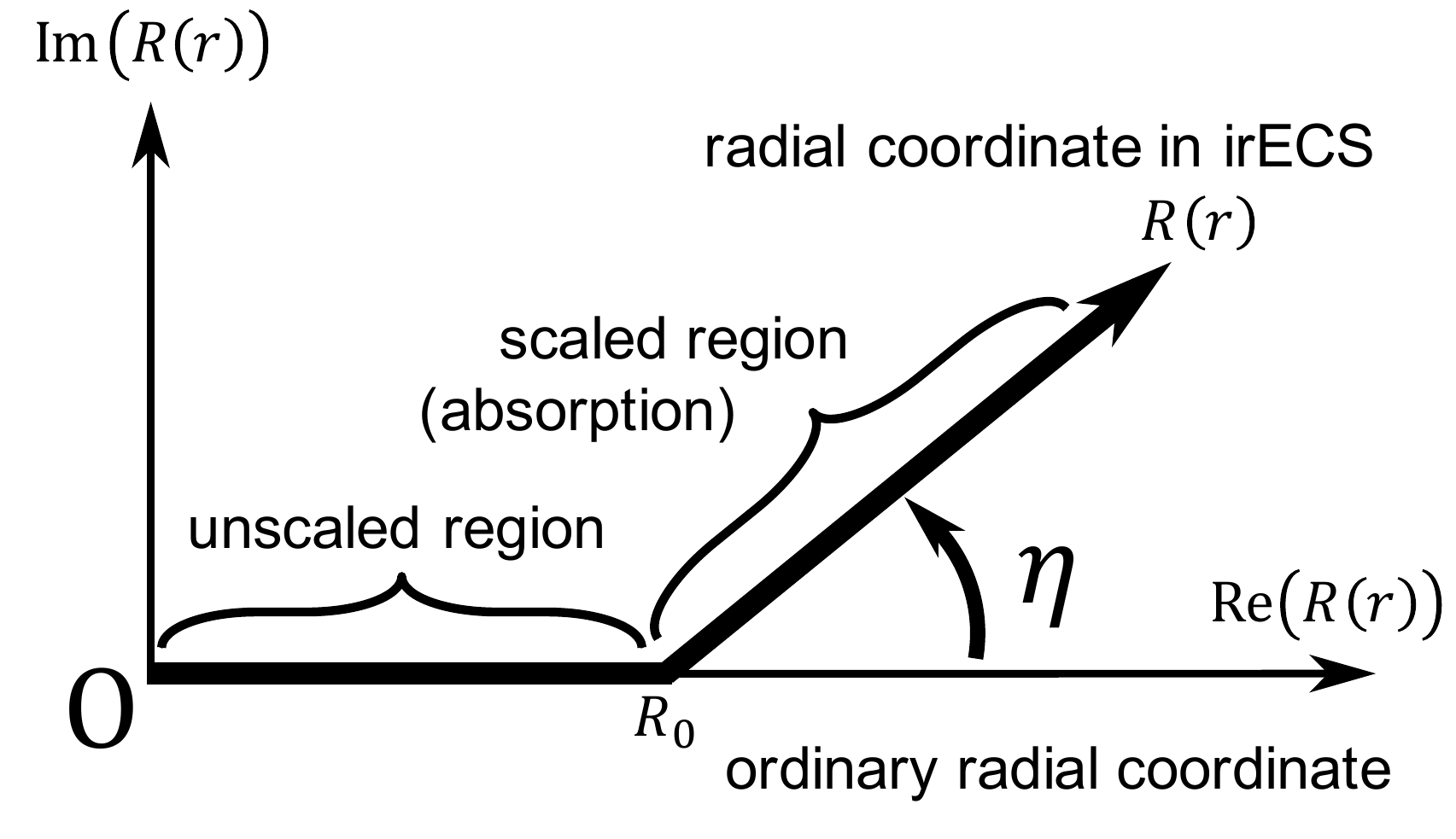}
	\caption{ Schematic illustration of radial exterior complex scaling contour $R(r)$ with scaling radius $R_0$ and scaling angle $\eta$.}
	\label{fig:ECScontour}
\end{figure}

Whereas our previous implementations \cite{Sato_2016,Sawada_2016} have used a mask function \cite{Krause_1992} and that another commonly used absorbing boundary is a complex absorbing potential~\cite{Riss_1996,Greenman_2010}, exterior complex scaling (ECS)~\cite{McCurdy_1991} is considered to be more sophisticated, which analytically continues the wave function into the complex plane (Fig.~\ref{fig:ECScontour}) without artificially modifying the system Hamiltonian nor the wave function. Furthermore,  the infinite range exterior complex scaling (irECS) method introduced in \cite{Scrinzi_2010} significantly improves the efficiency over standard ECS by using a exponentially damped basis, thus moving the reflecting boundary to infinity.

The application of ECS and irECS, originally formulated for single-electron problems, to strongly driven multielectron systems with the addition of the interelectronic Coulomb interaction has been so far limited.
McCurdy {\it et al.} \cite{McCurdy_1991} introduced ECS to two-electron systems where the Coulomb interaction was approximated in the radial limit. 
Haxton {\it et al.}  \cite{Haxton_2011} used ECS in their MCTDHF implementation but mainly dealt with photoionization rather than strong-field phenomena.
Telnov {\it et al.} \cite{Telnov_2013} applied ECS to the time-dependent density functional theory to simulate high-harmonic generation from Ar. In the scaled region, however, they neglected the laser field and replaced the time-dependent Hartree and exchange-correlation potentials with their initial values. 
Majety {\it et al} \cite{Majety_2015} have recently proposed the hybrid anti-symmetrized coupled channels method to calculate fully differential photoelectron spectra of multielectron systems subject to strong laser fields. Though irECS is used in the implementation, only an electronic coordinate is scaled in each channel as the method allows only single ionization.
Zielinski {\it et al.} \cite{Zielinski_2016} have applied irECS to two-electron systems, where the both electronic coordinates are scaled.
However, to our knowledge, irECS has never been applied to TD-MCSCF methods yet.

In this study, we introduce ECS and irECS to the TD-CASSCF method for \textit{ab initio} simulations of multielectron dynamics in atoms subject to intense laser fields. While minimally neglecting the Coulomb force from electrons residing in the scaled region, our implementation retains all the other nuclear-electron, electron-electron, and laser-electron interactions.
We achieve stable and highly accurate simulations of nonperturbative strong-field phenomena such as tunneling ionization and HHG with considerably reduced computational costs.

This paper is organized as follows. 
In Sec.~II, the TD-CASSCF method is briefly reviewed. 
In Sec.~III, we describe our numerical implementation of irECS, adopting the spherical finite-element discrete variable representation (FEDVR).
Section~IV discusses how to apply ECS and irECS to the TD-CASSCF method. 
In Sec.~V, Numerical examples are presented. 
Conclusions are given in Sec.~VI.
We use Hartree atomic units unless otherwise indicated.

\section{The TD-CASSCF method}
We consider the multielectron dynamics of an $N$-electron atom with atomic number $Z$ in a laser field $E(t)$ linearly polarized in the $z$ direction, described by the Hamiltonian:
\begin{gather}
	\label{eq:Ham}
	H(t) = \sum_{i=1}^{N} h(\vec{r}_i, \nabla_i, t) +  \sum_{i=1}^{N-1} \sum_{j=i+1}^{N} g(\vec{r}_i, \vec{r}_j),
\end{gather}
with,
\begin{gather}
	h(\vec{r}_i, \nabla_i, t) =  \left( - \frac{\nabla_i^2}{2} - \frac{Z}{r_i} - i \vec{A}(t) \cdot \nabla_i  \right), \\
	g(\vec{r}_i, \vec{r}_j) = \frac{1}{|\vec{r}_i - \vec{r}_j |},
\end{gather}
where $\vec{A}(t) =  - \int_{-\infty}^t \vec{E}(t') dt'$ denotes the vector potential. The velocity gauge is used, since ECS works only with it, and not with the length gauge \cite{McCurdy_1991}.

In the TD-CASSCF method \cite{Sato_2013}, the total wave function is given by,
\begin{equation}
	\Psi = \hat{A}\left[\Phi_{\rm fc}\Phi_{\rm dc}\sum_I\Phi_I C_I\right],
\end{equation}
where $\hat{A}$ denotes the antisymmetrization operator, $\Phi_{\rm fc}$ and $\Phi_{\rm dc}$ the closed-shell determinants formed with $n_{\rm fc}$ FC and $n_{\rm dc}$ DC orbitals, respectively, and $\{\Phi_I\}$ the determinants constructed from  $n_{\rm a}$ active orbitals. The active electrons are fully correlated among the active orbitals as in the MCTDHF method. 

The equations of motion (EOMs) that describe the temporal evolution of the CI coefficients $\{C_{I}\}$ and the orbital functions $\{\psi_p\}$ are derived by use of the time-dependent variational principle.
The EOMs have various forms depending on the choice of time derivatives of the orbitals \cite{Kato_2004, Caillat_2005, Sato_2013}.
In this paper, we choose the forms in our previous study \cite{Sato_2016}.
The EOMs for the CI coefficients are given by,
\begin{gather}
	i\frac{d}{dt} C_I(t) = \sum_J \braket{\Phi_J | \hat{g} |\Phi_I}C_J(t),
	\label{eq:CI-EOM}
\end{gather}
where $\hat{g}$ denotes the second-quantized expression of the second term in Eq.~(\ref{eq:Ham}).
The EOMs of the orbitals read,
\begin{gather}
\label{eq:eomorb}
	i \frac{d}{dt} \ket{\psi_p} = \hat{h} \ket{\psi_p} + \hat{Q} \hat{F}  \ket{\psi_p} + \sum_{q} \ket{\psi_q}  R^q_p,
\end{gather}
where $\hat{h}$ denotes the one-particle part [first term of Eq. \eqref{eq:Ham}] in second quantization, $\hat{Q} = 1 - \sum_{q} \Ket{\psi_q} \Bra{\psi_q}$ the projector onto the orthogonal complement of the occupied orbital space.
$\hat{F}$ is a non-local operator describing contribution from the interelectronic Coulomb interaction, defined as 
\begin{gather}
	\hat{F} \ket{\psi_p} = \sum_{oqsr} (D^{-1})^o_p P^{qs}_{or} \hat{W}^r_s \ket{\psi_q},
\end{gather}
where $D$ and $P$ are the one- and two-electron reduced density matrices, and $\hat{W}^r_s$ is the mean field potential, given, in the coordinate space, by
\begin{gather}
	W^{r}_{s} \left(\vec{r} \right) = \int d \vec{r}^\prime \frac{\psi_{r}^{*} (\vec{r}^\prime) \psi_{s} ( \vec{r}^\prime )}{| \vec{r} - \vec{r}^\prime | } .
	\label{eq:W}
\end{gather}
The matrix element $R^q_p$,
\begin{gather}
\label{eq:orbital-time-derivative}
	R^q_p = i \braket{\psi_q | \dot{\psi_p} } - h^q_p,
\end{gather}
where $h^q_p = \braket{\psi_q|\hat{h}|\psi_p}$, determines the components of the time derivative of the orbitals within the subspace that the occupied orbitals span.
The elements within one orbital subspace (core or active) can be arbitrary Hermitian matrix elements, and, in this paper, we set them to zero, i.e., $R^i_j = R^t_u = 0$, where $i$ and $j$ belong to the core orbital space and $t$ and $u$ belong to the active orbital space.
On the other hand, the elements between the core and active subspaces are given by, 
\begin{align}
	\label{eq:R_ti}
		&R^{t}_i = \left(R^i_t\right)^* =  {-h^t_i} - \vec{E}(t) \cdot \vec{r}^{\,t}_{\,i} \text{~~~~~~} (\text{for } i \in \text{frozen core}) \\
		\label{eq:R1}
		&R^{t}_i = \left(R^i_t\right)^*  =  \sum_u [(2-D)^{-1}]^{t}_{u} (2F^u_i - \sum_v D^u_v F^{i*}_v) \\
		&\text{~~~~~~}\text{~~~~~~} (\text{for } i \in \text{dynamical core}),  \notag
\end{align}
where $F^u_i = \braket{\psi_u | \hat{F} | \psi_i}$, and $\vec{r}^{\,t}_{\,i}$ denotes a matrix element of the position vector $\vec{r}$.
It should be noted that frozen core orbitals, which are time-independent in the length gauge, are to be propagated in the velocity gauge as \cite{Sato_2016},
\begin{equation}
\psi_i( \vec{r}, t)  =  
	 {\rm e}^{- i \vec{A}(t) \cdot \vec{r}} \psi_i( \vec{r}, 0) \qquad (\text{for } i \in \text{frozen core}).
\end{equation}

\section{Implementation of infinite-range ECS with FEDVR method}
\subsection{Exterior complex scaling for a single-electron system}

In this subsection, we briefly review exterior complex scaling for a single-electron system. Let us consider the velocity-gauge TDSE for a single-electron system in a laser field,
\begin{eqnarray}
	i \frac{\partial}{\partial t}\Psi(\vec{r}, t) &=& h_1(t)  \Psi(\vec{r}, t) \notag \\
	&=& \left(-\frac{\nabla^{2}}{2} +  V(\vec{r}) - i \vec{A}(t) \cdot \nabla \right) \Psi(\vec{r}, t),  \label{eq:single-electron TDSE}
\end{eqnarray}
with $V(\vec{r})$ being a system potential. 
For polar coordinated, ECS is based on the scaling
\begin{equation}
	\label{eq:rchange}
	r \rightarrow R(r) = 
	\begin{cases}
		r & (r < R_0) \\
		R_0 + (r-R_0) {\rm e}^{\lambda + i \eta} & (r > R_0),
	\end{cases}
\end{equation}
where $\lambda$ and ${\eta}$ is real numbers, and specifically $\eta$ is called a scaling angle.
For $\eta>0$, outgoing waves exponentially decay at radii $r>R_0$ and numerically vanish before they reach the simulation boundary and are unphysically reflected.

The transformation Eq.~(\ref{eq:rchange}) defines an ``exterior complex scaling operator" $U_{\eta R_0}$
\begin{equation}\label{eq:Ueta}
	(U_{\eta R_0} \Psi) (\vec{r}): = 
	\begin{cases}
		\Psi( \vec{R}(r)) & (r < R_0) \\
		{\rm e}^{\frac{\lambda + i \eta}{2}} \dfrac{R(r)}{r} \Psi( \vec{R}(r)) & (r > R_0),
	\end{cases}
\end{equation}
where,
\begin{equation}
	\vec{R}(r) = \frac{R(r)}{r}\vec{r}.
\end{equation}
The factor ${\rm e}^{\frac{\lambda + i \eta}{2}} R(r)/r$ ensures that $U_{\eta R_0}$ is unitary for $\eta=0$. 
In the unitary case, one can replace $h_1(t)$ in Eq.~(\ref{eq:single-electron TDSE}) with
\begin{equation}
\label{unitary}
h_{\eta=0 R_0}(t)=U_{\eta=0 R_0}h_1(t)U^{-1}_{\eta=0 R_0},
\end{equation}
without changing the dynamics. The solution for the scaled Hamiltonian is trivially $\Psi_{\eta=0 R_0}:=U_{\eta=0 R_0} \Psi$ and coincides with the unscaled solution $\Psi$ for $r<R_0$.

In the ECS case, the scaled operator is $h_{\eta R_0}(t)=h_1(t)$  on $r<R_0$ and for $r>R_0$ 
\begin{equation}\label{eq:hEta}
h_{\eta R_0} = -\frac{1}{2}\nabla_{\eta R_0}^2 + V[\vec{R}(r)] - i \vec{A}(t) \cdot \nabla_{\eta R_0},
\end{equation}
with the scaled nabla operator $\nabla_{\eta R_0}$ given by
\begin{eqnarray}
	\nabla_{\eta R_0}
	&=& \vec{e}_r \frac{1}{ {\rm e}^{\lambda + i \eta} r} \frac{\partial}{\partial r} r  \notag \\ 
	&+& \frac{1}{R(r) \sin \theta } \left(  \vec{e}_{\theta} \frac{\partial}{\partial \theta} \sin \theta  + \vec{e}_{\phi}  \frac{\partial}{\partial \phi} \right).
\end{eqnarray}
This form of the scaled operator is formally obtained by analytically continuing that of the unitary case [Eq.~\eqref{unitary}] with $\eta = 0 \rightarrow \eta \not = 0$ \cite{Scrinzi_1993}.
The essential point of ECS is that, given sufficient analyticity properties of $h_{\eta R_0}$, 
also for $\eta>0$ the solution $\Psi_{\eta R_0}$ remains invariant on $r<R_0$,
while it decays exponentially in the absorbing region \cite{Scrinzi_2010}.

On formal grounds one may expect such a behavior. However, it is not at all obvious as the operator $U_{\eta R_0}$ and its inverse 
are poorly defined for $\eta>0$.
This mathematical fact is reflected in numerical breakdown when approximating the inverse $U^{-1}_{\eta R_0}$ in
any discretization.

For the numerical solution of the complex scaled TDSE with the simple scaling of Eq.~(\ref{eq:hEta}) one needs to ensure that the discretization method can represent the 
discontinuous behavior of the solution at $r=R_c$, Eq.~(\ref{eq:Ueta}). This is case for the FEDVR basis set described below.

While ECS is usually applied on a finite discretization range, one can infinitely extend the scaled region by using a finite number of exponentially damped basis functions \cite{Scrinzi_2010}. This method, called infinite-range ECS, not only has a conceptual advantage of simulating the entire space with artificially modifying neither the system Hamiltonian nor the wave functions, but also has achieved high accuracy and efficiency with a considerably smaller number of basis functions \cite{Scrinzi_2010}.

\subsection{FEDVR basis for infinite-range ECS}


In this paper, we implement irECS with a spherical-FEDVR basis \cite{Rescigno_2000, McCurdy_2004}, which is used in our TD-CASSCF code \cite{Sato_2016}. 

Here, as usual, we set $\lambda=0$ in the scaling factor. 
For the discretization in radial direction we absorb the factor $R(r)$ into the basis functions $ry_i(r)$, 
which is equavalent to working with radial functions 
$\Phi(\vec{r})=r\Psi(\vec{r})$ and further simplifies the expression for $\nabla_{\eta R_0}$. 
For the implementation we follow Refs.~\cite{Rescigno_1997, McCurdy_2004}.

In FEDVR methods with a finite range, Gauss-Lobatto and Gauss-Radau quadrature points are usually used in each finite element for integral evaluation.
For irECS, instead, we introduce Gauss-Laguerre-Radau quadrature points \cite{Gautschi_2000,Weinmueller_2017} 
to construct DVR basis functions in the last finite element extending to infinity.
Gauss-Laguerre-Radau quadrature
approximates the semi-infinite integral of an exponentially damped function
as
\begin{gather}
\label{eq:GLR}
\int_{r_\text{L}}^{\infty} dr e^{-\alpha (r - r_L)} f(r) \approx \sum_{i = 1}^{N_\text{grid}} w_i f(r_i) \\
r_L = r_1 < r_2 < \cdots < r_{N_\text{grid} }\notag
\end{gather}
with $w_i$'s and $r_i$'s being weights and grid points, respectively. We choose the lower element boundary as $r_L=R_0$.
As discussed in Refs.~\cite{Rescigno_1997, McCurdy_2004}, the integration weights in complex scaled region $r\geq R_0$ 
are multplied by $e^{i\eta}$.

For irECS, one uses exponentially damped functions as the FEDVR basis functions on the last element,
\begin{gather}
	y_i(r) =
	\begin{cases}
		\displaystyle e^{- \frac{\alpha}{2} (r - r_L)} \frac{1}{r} \frac{ L_i(r)}{\sqrt{w_i}} & (r \ge r_L) \\
		0 & (r < r_L) \\
	\end{cases}
	\label{eq:irLagrange}
\end{gather}
%
%
with Lagrange polynomials,
\begin{gather}
L_i(r) = \displaystyle \prod_{j \not = i} \dfrac{r-r_j}{r_i - r_j}. 
\label{eq:Lagrange}
\end{gather}
Note that these basis functions are not truncated within a finite range unlike usual FEDVR basis, but extend to the infinite range and decay exponentially due to 
the factor $e^{- \frac{\alpha}{2} (r - r_L)}$.
This infinitely-extended exponential tail can describe exponentially damped orbital functions by ECS and provides high accuracy with a small number of basis functions.

The basis functions appear as orthonormal under the approximate Gauss quadrature,
\begin{align}
	\int_{0}^{\infty} dr \, r^2  y_i(r) y_j(r) &\approx \sum_{k = 1}^{N_\text{grid}} w_k e^{\alpha (r_k - r_L)} y_i(r_k)^* y_j(r_k) \notag \\
	&= \delta_{ij}.
\end{align}
Thus, in the last finite element a radial part of scaled orbital functions $\varphi(r)$ is expressed by
\begin{gather} 
	\varphi(r) \approx \sum_i^{N_\text{grid}} c_i y_i(r)\\
	c_i = \int_0^{\infty} dr \, r^2 y_i(r) \varphi(r) \approx \sqrt{w_i} r_i \varphi(r_i).
\end{gather}
Likewise, the matrix elements of one-body potentials are diagonal,
\begin{equation}
	V_{ij} = \int_{0}^{\infty} dr \, r^2 y_i(r) V(r) y_j(r) \approx V(r_i) \delta_{ij}.
\end{equation}

The first derivative of the FEDVR basis functions are given by
\begin{gather}
	\frac{\partial}{\partial r} y_i(r)  =
	 -\frac{1}{r} y_i(r)  + \frac{1}{\sqrt{w_i}} \dfrac{e^{-\frac{1}{2} \alpha (r - r_L)}}{r} P_i(r), 
\end{gather}
where
\begin{align}
	 P_i(r) 
	 &=  - \dfrac{\alpha}{2}  L_i(r) + \dfrac{\partial}{\partial r}L_i(r) \notag \\
	 &= \begin{cases}
		\dfrac{1}{r_i - r_j} \displaystyle \prod_{k \not = i, j} \dfrac{r_j - r_k}{r_i - r_k} & \text{for } r=r_j \text{, } i \not= j \\[12pt]
		- \dfrac{1}{2 w_1} \delta_{i 1} & \text{for }  r=r_j \text{, } i = j.
	\end{cases} \label{eq:firstD}
\end{align}
Thus, the matrix elements of the radial second derivative operator can be expressed under Gauss quadrature by using a partial integral,
\begin{align}
	\int_{0}^{\infty} r^2 dr \, y_i(r) \frac{1}{r} \frac{\partial^2}{\partial r^2} r y_j(r) &= - \int_{0}^{\infty} dr \frac{\partial}{\partial r} ( r y_i(r))  \frac{\partial}{\partial r} (r y_j(r)) \notag \\
	&\approx - \sum_{k} \frac{w_k}{\sqrt{w_i w_j}} P_i(r_k) P_j(r_k)
\end{align}

For simplicity, we have discussed without considering the bridge function to connect the element boundary between the last element and the second to last element. In the actual implementation, we introduced this as well as in the usual FEDVR method \cite{Rescigno_2000}.

\section{application of ECS to the TD-CASSCF multielectron dynamics}
\label{sec:application of ECS}

In this Section, we discuss how to apply ECS to TD-CASSCF simulations of the multielectron dynamics involving the interelectronic Coulomb interaction.
By analogy with the single-electron case, we propagate the scaled orbital function $\hat{U}_{\eta R_0} \Ket{ \psi_p}$ rather than the unscaled $\Ket{ \psi_p}$, by transforming Eq.~\eqref{eq:eomorb} into the scaled EOMs of the orbitals,
\begin{multline}
\label{eq:ecseomorb}
i  \frac{d}{dt} (\hat{U}_{\eta R_0} \Ket{ \psi_p})  = (\hat{U}_{\eta R_0} \hat{h} \hat{U}_{\eta R_0}^{-1}) (\hat{U}_{\eta R_0} \Ket{\psi_{p}}) \\
+ \left[ 1 - \sum_{t}  (\hat{U}_{\eta R_0} \Ket{\psi_t}) (\Bra{\psi_t}  \hat{U}_{\eta R_0}^{-1}) \right]  \\
\times \sum_{oqsr} ( D^{-1})^o_p P^{qs}_{or}  (\hat{U}_{\eta R_0} \hat{W}^r_s \hat{U}_{\eta R_0}^{-1})  (\hat{U}_{\eta R_0} \Ket{\psi_q}) \\
 + \sum_{q} (\hat{U}_{\eta R_0} \Ket{\psi_q}) R_{p}^{q}.
\end{multline}
A significant difference from the EOMs without ECS is that $\{ \Bra{\psi_p}\hat{U}_{\eta R_0}^{-1} \}$is required, instead of $\{ \Bra{\psi_p} \}$, to apply $\hat{Q} = 1 - \sum_t \ket{\psi_t}\bra{\psi_t} $ and evaluate $W$ and $R$. It is formally defined in the coordinate space as
%
{\begin{equation}
	(\Bra{\psi_q} \hat{U}_{\eta R_0}^{-1}) \Ket{\vec{r}} = 
	\left[ \Bra{\vec{r}} 
		\left( 
			\hat{U}_{(-\eta) R_0} \Ket{\psi_q}
		\right)
	\right]^*.
\end{equation}}

%

It should be noticed that information of $\{ \Bra{\psi_p}\hat{U}_{\eta R_0}^{-1} \}$  is available in the unscaled region but \emph{not} available in the scaled region during the simulation, which poses a problem. 
Although formally one might attempt to obtain these by analytically
continuing $\{\hat{U}_{\eta R_0}\Ket{\psi_p}\}$ , such a procedure turns out to be numerically
unstable.

Since the scaled region is usually far from the origin, it is reasonable to assume that the scaled part of the orbital functions hardly affects the electron dynamics close to the nucleus and that the interaction between electrons residing in the scaled region is negligible.
Thus, we approximately neglect $\{ \Bra{\psi_p}\hat{U}_{\eta R_0}^{-1} \}$ in the scaled region wherever their information is necessary to evaluate the right-hand side (RHS) of Eq.~\eqref{eq:ecseomorb}.

Specifically, the scaled mean field operator is approximated as,
\begin{align}
	U_{\eta R_0} W^r_s (\vec{r}) U_{\eta R_0}^{-1} 
	&= W_s^r(\vec{R}(r)) \notag \\
	&\approx \int_{r^\prime < R_0} \!\!\!\!\!\! d\vec{r}^\prime \frac{ \psi_{r}^* (\vec{r}^\prime) \psi_{s} ( \vec{r}^\prime )}{| \vec{R}(r) - \vec{r}^\prime | }  \notag\\
	&\equiv {W'}^{r}_{s}(\vec{R}(r)) \label{eq:Wprime}
\end{align}
Here, it should be noticed that the Coulomb force acting on a scaled-region electron ($r > R_0$) from an unscaled-region electron ($r^\prime < R_0$) is not neglected. Hence, the effect of the ionic Coulomb potential is properly taken into account in the dynamics of departing electrons.
The way to numerically evaluate the truncated scaled mean field operator $W{'}^{r}_{s} (\vec{R}(r))$ is given in Appendix \ref{sec:append1}.
Then, in the second term of the RHS of Eq.~\eqref{eq:ecseomorb},
\begin{align}
	& (\Bra{\psi_t} \hat{U}_{\eta R_0}^{-1}) ( D^{-1})^o_p P^{qs}_{or} (\hat{U}_{\eta R_0} \hat{W}^{r}_{s}   \hat{U}_{\eta R_0}^{-1})  (\hat{U}_{\eta R_0} \Ket{\psi_q}),
\end{align}
is approximated as,
\begin{equation}
	\label{eq:QW_approx}
	( D^{-1})^o_p P^{qs}_{or} \int_{r < R_0} \!\!\!\!\!\! d\vec{r} \psi_{t}^* (\vec{r})  {W'}^{r}_{s} (\vec{R}(r))  \psi_{q} ( \vec{r} ).
\end{equation}
Similarly, in the evaluation of Eq.~\eqref{eq:R_ti}, the RHS is approximated as,
\begin{align}
	\label{eq:h1_approx}
	h_i^t \approx \int_{r < R_0} \!\!\!\!\!\! d\vec{r} \psi_{t}^* (\vec{r}) h(\vec{r}) \psi_{i}(\vec{r}).
\end{align}
However, since $\{\psi_{i}(\vec{r})\}$ in Eq.~\eqref{eq:R_ti} is a frozen core orbital, which is fixed at an initial bound state and exponentially decays with the distance from the origin increasing, the truncation in Eq.~\eqref{eq:h1_approx} gives almost no error. 
In order to evaluate the matrix elements of $F$ in the RHS of Eq. \eqref{eq:R1} and to propagate CI coefficients using Eq.~\eqref{eq:CI-EOM}, we need to evaluate the following Coulomb matrix elements,
\begin{align}
	W^{rp}_{sq} = \int d\vec{r} d\vec{r}^\prime \frac{\psi_{r}^{*} (\vec{r}) \psi_{p}^{*} (\vec{r}^\prime) \psi_{q} ( \vec{r}^\prime ) \psi_{s} ( \vec{r} )}{| \vec{r} - \vec{r}^\prime | },
\end{align}
which we approximate as, truncating the integral within the unscaled region as well as Eq.~\eqref{eq:QW_approx},
\begin{align}
	\label{eq:CI_approx}
	W^{rp}_{sq} \approx  \int_{r < R_0} \!\!\!\!\!\! d\vec{r} \psi_{p}^* (\vec{r})  {W'}^{r}_{s} (\vec{R}(r))  \psi_{q} ( \vec{r} ).
\end{align}
The validity of these truncation procedures will be numerically assessed below in Sec.~\ref{sec:result}. 
%

\section{Numerical results}
\label{sec:result}
In this section, we assess performance of the implementation of irECS to the TD-CASSCF method described in the previous section, simulating many-electron atoms in an intense near-infrared laser pulse.
We assume a laser field linearly polarized in the $z$ direction of the following form:
\begin{equation}
	E(t) = \sqrt{I_0} \sin \omega t \sin^2 \left( \pi \frac{t}{NT} \right) ,\  ( 0 \leq t \leq NT),
	\label{eq:laser}
\end{equation}
where $I_0$ is a peak intensity, $T$ is a period at the central frequency $\omega = 2\pi/T$ and $N$ is the total number of optical cycles.
We gauge the performance of simulations with irECS against nominally ``exact" results converged with respect to a simulation box size and obtained with the mask function boundary.
In the latter, orbital functions are multiplied by a mask function,
\begin{equation}
	M(\vec{r}) = 
	\begin{cases}
		1 & \text{for } |\vec{r}| < R_\text{mask} \\
		\cos^{\frac{1}{4}} \left(\dfrac{\pi}{2} \dfrac{|\vec{r}| - R_\text{mask}}{R_\text{max} - R_\text{mask}} \right) & \text{for } |\vec{r}| \geq R_\text{mask},
	\end{cases}
\end{equation}
after each time step, where $R_\text{mask}$ and $R_\text{max}$ denote the absorption boundary and the simulation box radius, respectively.
%
%

\subsection{Helium}
We first simulate a helium atom subject to a laser field of $8.0 \times 10^{14} \text{ W/m$^2$}$ peak intensity, $\lambda = 800$ nm wavelength, and five-optical-cycle foot-to-foot pulse duration ($N = 5$). 
We use five active orbitals, each expanded with 47 spherical harmonics.
For the radial direction, each finite element has 21 grid points and is 4 a.u.\,long, except for the last irECS element, which extends to infinity. 
The number of the grids in the irECS element is same as other finite elements.
We performed simulations with four different absorption-boundary conditions, as listed in Table. \ref{tab:detailHe}.
The scaling angle $\eta$ is fixed to $15^\circ$.

\begin{table}[tb]
	\caption{Absorbing boundaries tested for He. $n_{\text{ua}}$ ($n_{\text{a}}$) denotes the number of grid points in the non-absorption (absorption) region, and $L_\text{a}$ the radial thickness of the absorption region. The radius $R_\text{max}$ of the whole simulation region is given by $R_\text{mask} + L_\text{a}$ or $R_{0} + L_\text{a}$.}
	\begin{ruledtabular}
		\begin{tabular}{cccccc}
			label & absorber & $R_{\text{mask}}$ or $R_0$ & $n_\text{ua}$ & $L_{\text{a}}$ & $n_\text{a}$\\
			\hline
			A  & mask   & 320 & 1600 & 80 & 400 \\
			B  & irECS  & 64 & 320 & $\infty$ & 40 \\
			C  & mask   & 64 & 320 & 8 & 40 \\
			D  & mask   & 124 & 620 & 40 & 200 \\
		\end{tabular}
	\end{ruledtabular}
	\label{tab:detailHe}%
\end{table} 

\begin{figure}[t]
      \begin{minipage}[t]{1\hsize}
      	\includegraphics[scale=0.8, clip, bb=0 0 288 144]{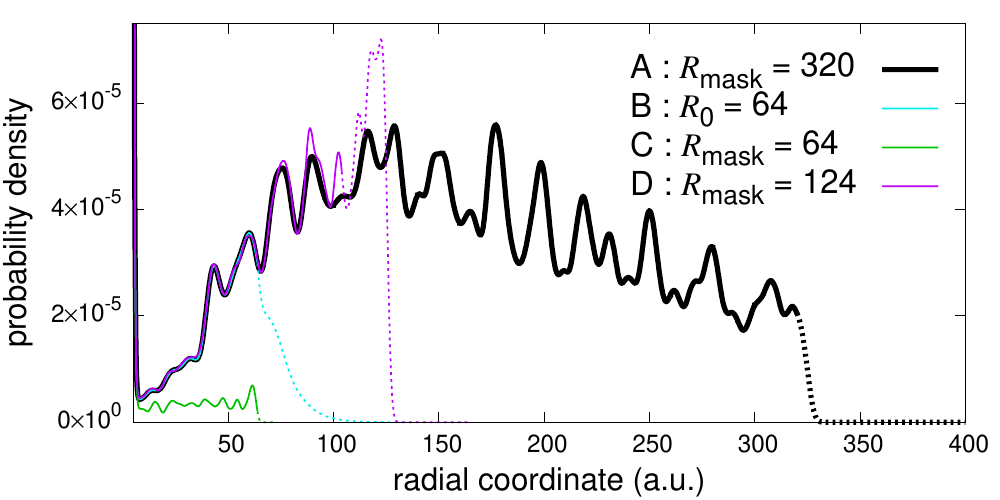}
      	\caption{Electron radial distribution function $\rho (r)$ after the laser pulse for the case of He exposed to a laser pulse with 800 nm wavelength and $8.0 \times 10^{14} \text{ W/m$^2$}$ peak intensity, calculated with different absorbing boundaries listed in Table \ref{tab:detailHe}.}
      	\label{fig:radHeLarge}
      \end{minipage} \\
      \begin{minipage}[t]{1\hsize}
      	\includegraphics[scale=0.8, clip, bb=0 0 288 144]{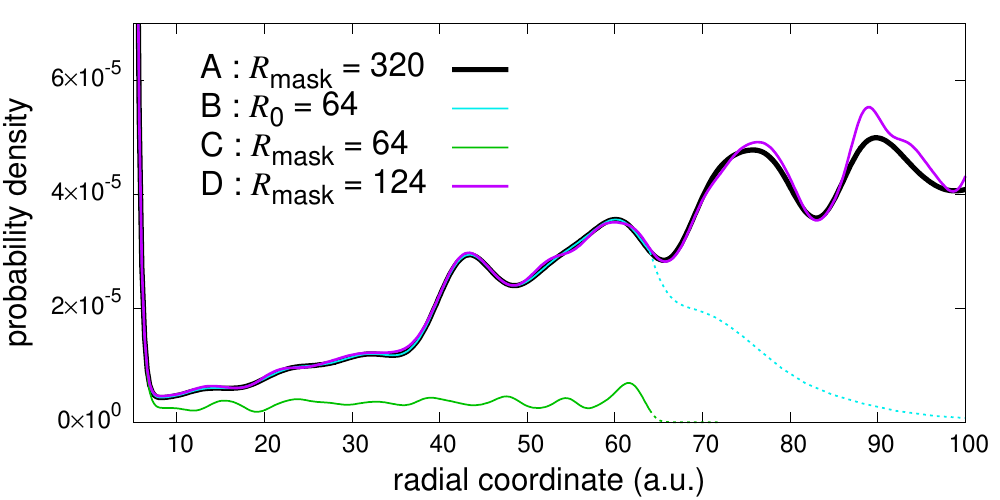}
	\caption{Enlarged view of Fig. \ref{fig:radHeLarge}.}
	\label{fig:radHe}
      \end{minipage}
      \begin{minipage}[t]{1\hsize}
      	\includegraphics[scale=0.8, clip, bb=0 0 288 144]{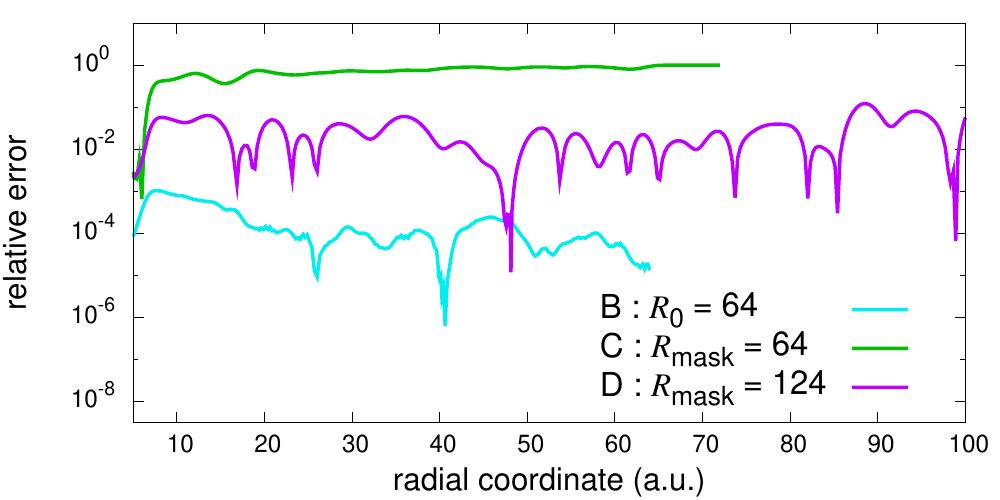}
	\caption{Relative errors of the radial distribution functions shown in Fig.~\ref{fig:radHe} compared to the result A.}
	\label{fig:radHeError}
      \end{minipage}
\end{figure}

\begin{figure}[t]
	\includegraphics[scale=0.8, clip, bb=0 0 288 144]{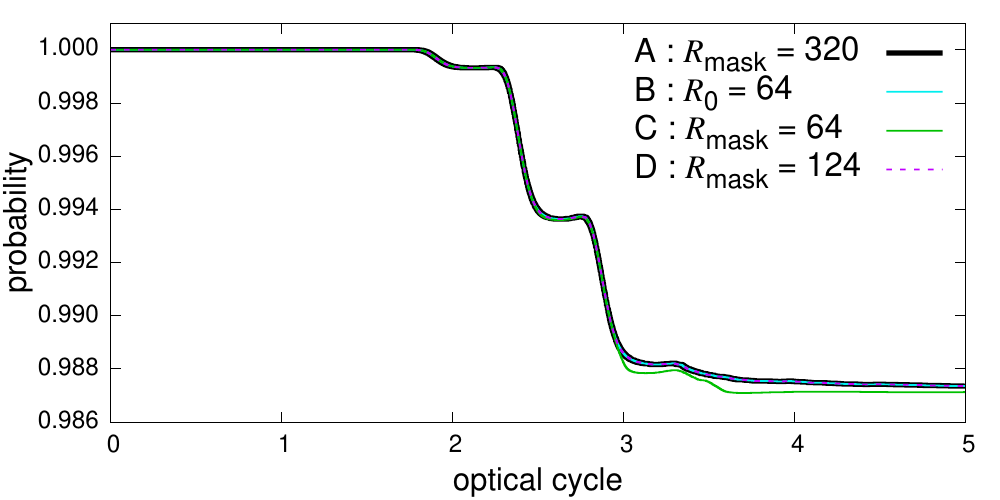}
	\caption{Time evolution of the survival probability, i.e., the probability of finding both electrons in He within 20 a.u.~radius, calculated with different absorbing boundaries listed in Table \ref{tab:detailHe}.}
	\label{fig:ipxHe}
\end{figure}

\begin{figure}[t]
	\includegraphics[scale=0.8, clip, bb=0 0 288 144]{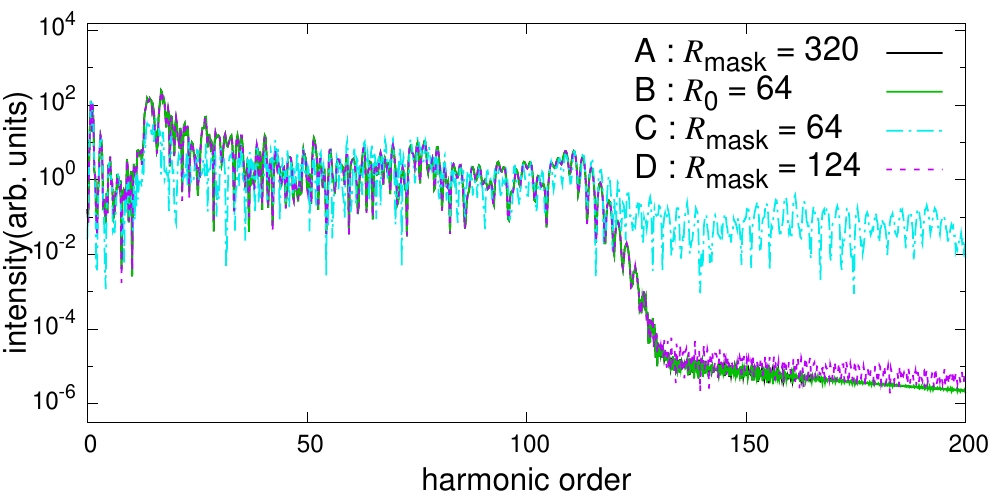}
	\caption{High harmonic spectra from He exposed to a laser pulse with 800 nm wavelength and $8.0 \times 10^{14} \text{ W/m$^2$}$ peak intensity, calculated with different absorbing boundaries listed in Table \ref{tab:detailHe}.}
	\label{fig:hhgHe}
\end{figure}%

Figures \ref{fig:radHeLarge} and \ref{fig:radHe} show the electron radial distribution function defined as,
\begin{equation}
	\label{eq:RDF}
	\rho(r) =  N {r}^2  \int d\sigma d\Omega d\vec{x}_2 d\vec{x}_3 \cdots \vec{x}_n \left|\Psi(\vec{x}, \vec{x}_2, \cdots, \vec{x}_n ) \right|^2,
\end{equation}
after the laser pulse, and Fig.~\ref{fig:radHeError} shows relative errors $|\rho(r) - \rho_{\text{A}}(r)|/\rho_{\text{A}}(r)$ compared to the radial distribution $\rho_{\text{A}}(r)$ for condition A.
Whereas the results (C and D) with the mask function deviate from the exact result (A, black thick solid lines), irECS (B) delivers the result with orders of magnitude smaller errors within the scaling radius $R_0$.
Figure \ref{fig:ipxHe} shows the temporal evolution of the probability of finding both electrons within the 20 a.u.~radius, which serves as a measure of survival probability.
The results with irECS (B) and the mask function with the larger absorption radius (D) agrees very well with the exact one (A), while that with the mask function (C) whose absorption radius $R_{\rm mask}$ is equal to $R_0$ deviates from the exact one after three optical cycles.
The degree of ionization is less than 1.5 \%, thus, the neglect of the Coulomb interaction in and from the scaled region (Sec.~\ref{sec:application of ECS}) leads to virtually no errors.

Figure \ref{fig:hhgHe} displays high-harmonic spectra, calculated as the magnitude squared of the Fourier transform of dipole acceleration. Again, the mask function with $R_{\rm mask}=64$ fails to reproduce, in particular, the sharp drop of the spectral intensity after the cutoff because of unphysical reflection, and $R_{\rm mask}=128$ is required for convergence.
On the other hand, the spectrum with irECS with $R_0=64$ shows excellent agreement with the exact one.

\subsection{Beryllium}
We move on to Beryllium with its ionization potential (9.3 eV) significantly smaller than that of He (24.6 eV), thus, much easier to ionize. 
We use $I_0 = 3.0 \times 10^{14} \text{ W/m$^2$}$, $\lambda = 800 {\rm nm}$ (the quiver radius is  28.5 a.u.), $N = 5$, and $(n_{\rm a},n_{\rm dc},n_{\rm fc}) = (4,0,1)$. 
As in the case of He, each orbital function is expanded with 47 spherical harmonics and discretized with radial finite elements 4 a.u.\ long except for the last irECS element. Each finite element, including the irECS element, has 21 grid points.
The scaling angle $\eta$ is set to be $15^\circ$.
Five different conditions used for absorption boundaries are listed on Table \ref{tab:detailBe}.

\begin{table}[tb]
	\caption{Absorbing boundaries tested for Be.}
	\begin{ruledtabular}
		\begin{tabular}{lccccr}
			 & absorber & $R_{\text{mask}}$ or $R_0$ & $n_\text{ua}$ & $L_{\text{a}}$ & $n_\text{a}$\\
			\hline
			A & mask & 320 & 1600 & 80 & 400 \\
			B & irECS & 40 & 200 & $ \infty $ & 40 \\
			C & irECS & 52 & 260 & $ \infty $ & 40 \\
			D & mask & 52 & 260 & $ 8 $ & 40 \\
			E & mask & 88 & 440 & $ 56 $ & 280 \\

		\end{tabular}
	\end{ruledtabular}
	\label{tab:detailBe}%
\end{table} 

\begin{figure}[t]
	\includegraphics[scale=0.8, clip, bb=0 0 288 144]{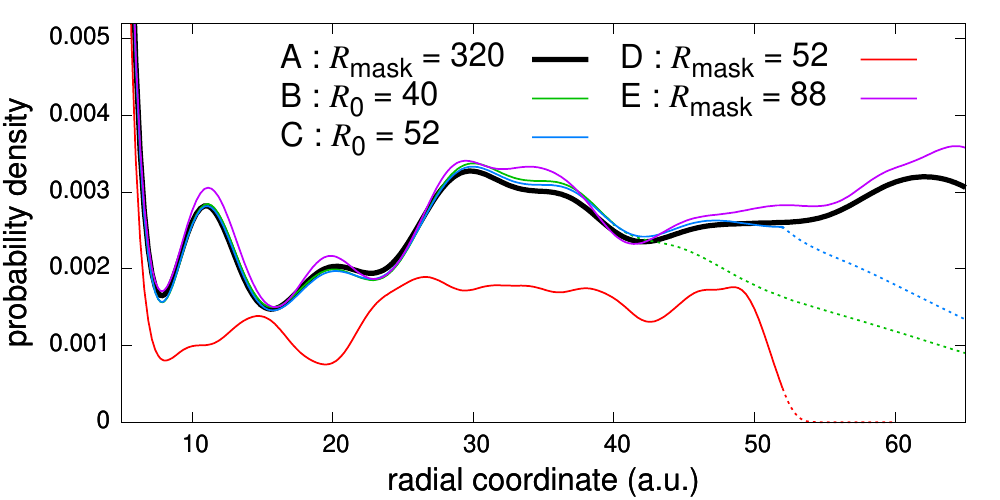}
	\caption{Electron radial distribution function $\rho (r)$ after the laser pulse for the case of Be exposed to a laser pulse with 800 nm wavelength and $3.0 \times 10^{14} \text{ W/m$^2$}$ peak intensity, calculated with different absorbing boundaries listed in Table \ref{tab:detailBe}.}
	\label{fig:radBe}
\end{figure}%

\begin{figure}[t]
	\includegraphics[scale=0.8, clip, bb=0 0 288 144]{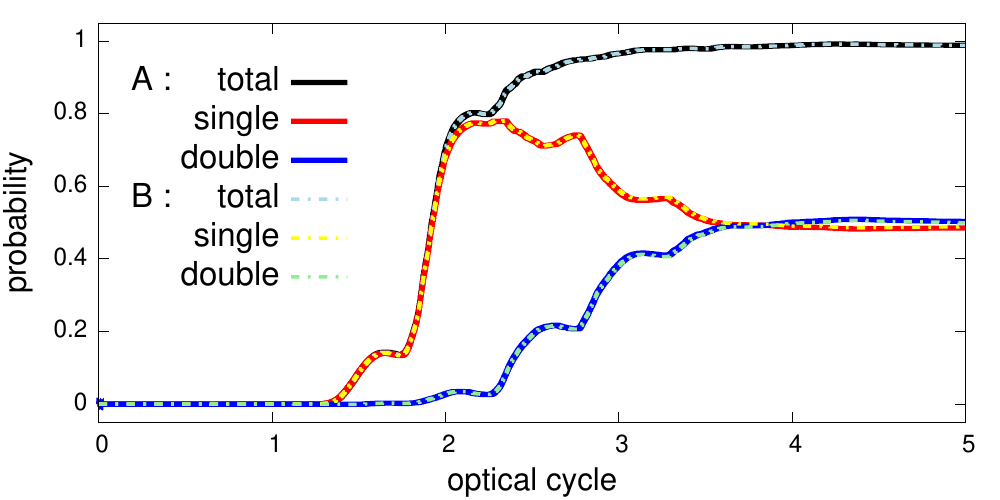}
	\caption{Time evolution of single, double, and total ionization probabilities of Be exposed to a laser pulse with 800 nm wavelength and $3.0 \times 10^{14} \text{ W/m$^2$}$ peak intensity. For convenience, we define single (double) ionization probability as that of finding one (two) electron(s) outside the 20 a.u.~radius. The total ionization probability is calculated as their sum.}
	\label{fig:ipxBe}
\end{figure}%

\begin{figure}[t]
	\includegraphics[scale=0.8, clip, bb=0 0 288 144]{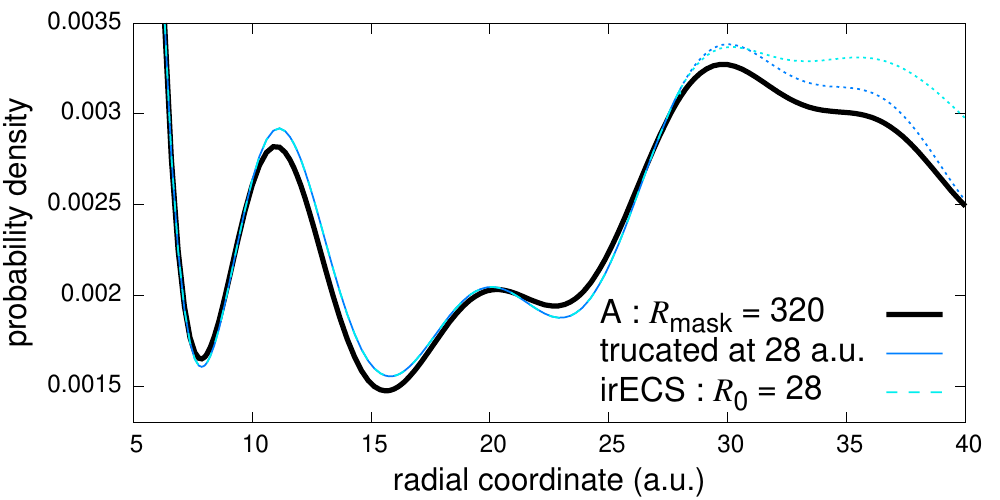}
	\caption{Electron radial distribution function $\rho (r)$ after the laser pulse for the case of Be exposed to a laser pulse with 800 nm wavelength and $3.0 \times 10^{14} \text{ W/m$^2$}$ peak intensity. We compare the results using the mask boundary ($R_{\rm mask}=320\,{\rm a.u.}$) without (thick solid) and with (thin solid) the integral truncations at 28 a.u., as described in Sec.~\ref{sec:application of ECS}, and the result using the irECS with $R_0=28\,{\rm a.u.}$ (dashed).}
	\label{fig:nodiffBe}
\end{figure}%

\begin{figure}[t]
	\includegraphics[scale=0.8, clip, bb=0 0 288 144]{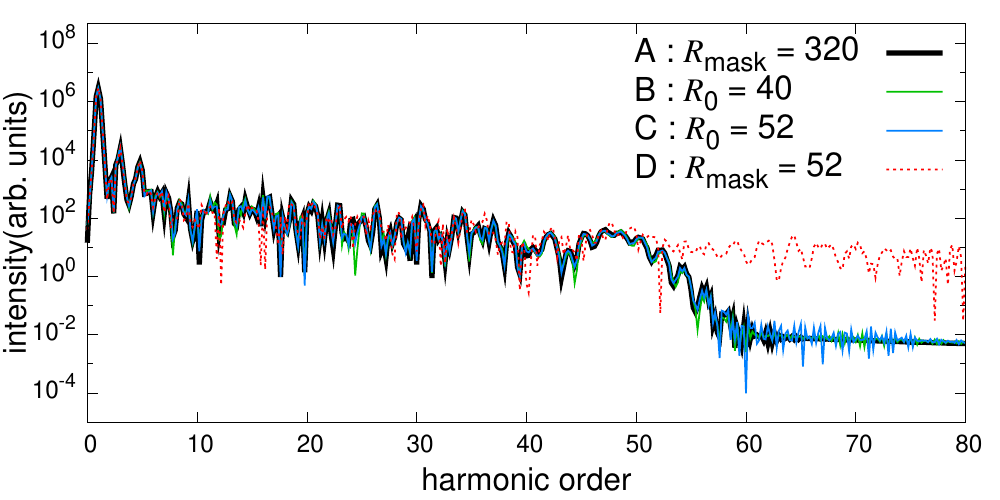}
	\caption{High harmonic spectra from Be exposed to a laser pulse with 800 nm wavelength and $3.0 \times 10^{14} \text{ W/m$^2$}$ peak intensity, calculated with different absorbing boundaries A-D listed in Table \ref{tab:detailBe}.}
	\label{fig:hhgBe}
\end{figure}%

Figure \ref{fig:radBe} compares the electron radial distribution functions after the pulse, calculated with different absorption boundaries. 
The irECS delivers much better results (B and C) than the mask function (D).
Nevertheless the irECS results slightly deviate from the exact solution (A) even if the scaling radius is almost twice the quiver radius.
In the present case, the Be atom is nearly totally ionized, and double ionization amounts to 50 \% (Fig. \ref{fig:ipxBe}). Hence, the deviation may be attributed to the neglect of the Coulomb interaction in and from the scaled region and/or the loss of information on the wave function in the scaled region. 
 
In order to reveal the effect of the latter, we have performed a simulation with a sufficiently large domain with the mask function ($R_\text{mask} = 320$ a.u. and  $R_\text{max} = 400$ a.u.) but with the integrals truncated at $r=28\,{\rm a.u.}$, as described in Sec.~\ref{sec:application of ECS}. 
We compare the result with the exact one and that from irECS with $R_0 = 28\,{\rm a.u.}$ in Fig.~\ref{fig:nodiffBe} . 
The ``truncated" and irECS results overlap each other and slightly deviate from the exact solution, which indicates that the slight deviations in Figs.~\ref{fig:radBe} and \ref{fig:nodiffBe} stem from the neglect of the Coulomb interaction in and from the scaled region. 
One may be surprised that the loss of information on orbital functions at the absorption boundary hardly affects simulation results within the absorption radius. 
This may be because the TD-CASSCF (and MCTHDF) equations of motion are derived by assuming the orthonormality of orbital functions, whether each of them may be (totally or partially) absorbed or not. Consequently, information on the absorbed part, though its explicit form is unknown, is partially retained, which enables accurate simulations.   
It should also be noticed that, since we construct the total wave function based on single-electron orbitals, even if one or more electrons are absorbed, we can continue to follow the associated dynamics of the other unabsorbed electrons. This is in great contrast to the time-dependent close-coupling simulations \cite{Ishikawa_2015, Parker_1996, Colgan_2004, Feist_2008}, where, if one electron reaches the absorption boundary, the corresponding part of the total wave function is completely lost.

In spite of the small discrepancy in Fig.~\ref{fig:radBe}, irECS gives the time evolution of single/double ionization (Fig.~\ref{fig:ipxBe}) and the high-harmominic spectrum (Fig.~\ref{fig:hhgBe}) in excellent agreement with the exact results. It is remarkable that the neglect of the Coulomb interaction in and from the scaled region is a good approximation and that irECS works excellently even under such massive double ionization. 
We have reduced computational costs by 66\% compared with the best case of the mask function (E) to obtain a converged high harmonic spectrum (B).

\subsection{Neon}
Finally, as a typical target atom used for attosecond-pulse generation, we simulate HHG from a Neon atom subject to a laser pulse with $I_0 = 8.0 \times 10^{14} \text{ W/m$^2$}$, $\lambda = 800\,{\rm nm}$, and $N = 3$.
We use 8 active orbitals and 1 dynamical core, i.e., $(n_{\rm a},n_{\rm dc},n_{\rm fc}) = (8,1,0)$. 
Each orbital function is expanded with 47 spherical harmonics and discretized with radial finite elements 4 a.u.\ long except for the last irECS element. Each finite element, including the irECS element, has 21 grid points.
The scaling angle $\eta$ is fixed to $15^\circ$. Three different conditions used for absorption boundaries are listed in Table \ref{tab:detailNe}.

\begin{table}[tb]
	\caption{Absorbing boundaries tested for Be.}
	\begin{ruledtabular}
		\begin{tabular}{lccccr}
			 & absorber & $R_{\text{mask}}$ or $R_0$ & $n_\text{ua}$ & $L_{\text{a}}$ & $n_\text{a}$\\
			\hline
			A & mask & 256 & 1280 & 64 & 320 \\
			B & irECS & 60 & 300 & $ \infty $ & 60 \\
			C & mask & 60 & 300 & $ 12 $ & 60 \\
		\end{tabular}
	\end{ruledtabular}
	\label{tab:detailNe}%
\end{table} 

\begin{figure}[t]
	\includegraphics[scale=0.8, clip, bb=0 0 288 144]{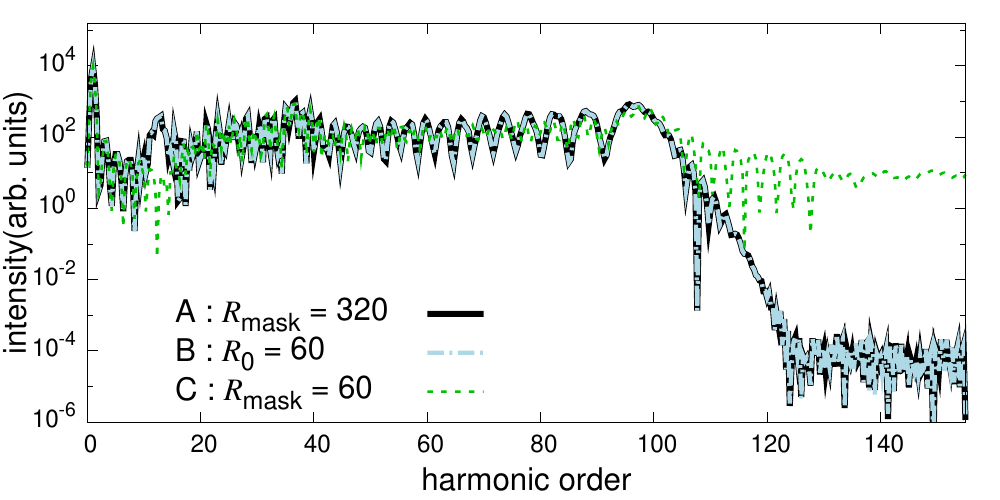}
	\caption{High harmonic spectra from Ne exposed to a laser pulse with 800 nm wavelength and $8.0 \times 10^{14} \text{ W/m$^2$}$ peak intensity, calculated with different absorbing boundaries listed in Table \ref{tab:detailNe}.}
	\label{fig:hhgNe}
\end{figure}%

If we use the same radius $R_0, R_{\rm mask}=60\,{\rm a.u.}$ of the non-absorbing region and number $n_a=60$ of grid points in the absorption region, the irECS result (B) perfectly overlaps with the ``exact" result (C) obtained with a large simulation box ($R_{\rm mask}=256\,{\rm a.u.}$), while the mask boundary (C) fails (Fig.~\ref{fig:hhgNe}). 
As in the case of He, the ionization probability (about 4~\%) is relatively small due to the large ionization potential (21.6~eV) of Ne, so that the truncation of integrals introduced to apply ECS to the TD-CASSCF method leads to almost no error.
The computational cost of the irECS simulation B is reduced by 80\% compared with case A.

\section{summary}

We have presented a successful numerical implementation of irECS as an efficient absorbing boundary to the TD-CASSCF method for multielectron atoms in intense laser fields.
It minimally neglects only the Coulomb force between electrons in the unscaled region and that acting from electrons in the scale region on those in the unscaled region.
For discretization of the scaled region, we have introduced Gauss-Laguerre-Radau quadrature points to construct exponentially dumped infinite-range FEDVR basis functions, thereby retaining their useful properties such as orthonormality and finiteness only at a grid point associated to each basis function.

We have applied the present method to He, Be, and Ne atoms, and calculated ionization probabilities and HHG spectra for intense near-infrared laser pulses. We have obtained   the results that perfectly agree with the converged results using much larger absorbing radii, even when atoms are massively ionized.
The demonstrated high accuracy indicates that the above-mentioned neglect of the Coulomb interaction, i.e., the truncation of integrals Eqs.~\eqref{eq:Wprime}, \eqref{eq:QW_approx}, \eqref{eq:h1_approx}, and \eqref{eq:CI_approx} is a good approximation, not significantly modifying simulated processes. 
Decreasing the size of the simulation box thanks to irECS has led to significant reduction of computational costs; by 66\% for Be and 80\% for Ne in the present case.

The present implementation for atoms uses the polar coordinate system with the spherical harmonics expansion, thus, its computational cost linearly scales with the radius of the simulation region.
If irECS is applied to our previously presented simulation code with three-dimensional Cartesian coordinates \cite{Sawada_2016}, even more efficiency gain is expected, which enables simulations of larger molecules and longer term simulations.

\begin{acknowledgments}
This research was supported in part by a Grant-in-Aid for Scientific Research (Grants No.~25286064, No.~26390076, No.~26600111, and No.~16H03881) from the Ministry of Education, Culture, Sports, Science and Technology (MEXT) of Japan and also by the Photon Frontier Network Program of MEXT.
This research was also partially supported by the Center of Innovation Program from the Japan Science and Technology Agency, JST, and by CREST (Grant No.~JPMJCR15N1), JST.
Y.O. gratefully acknowledges support from the Graduate School of Engineering, The University of Tokyo, Doctoral Student Special Incentives Program (SEUT Fellowship).
\end{acknowledgments}

\appendix
\section{Scaled interelectronic Coulomb interaction}
\label{sec:append1}
This Appendix briefly describes how to numerically evaluate ${W'}^r_s (\vec{R}(r))$ [Eq.~(\ref{eq:Wprime})].
By using the multipole expansion of $1/|\vec{r} - \vec{r}^\prime|$, 
\begin{equation}
	\frac{1}{|\vec{r} - \vec{r}^\prime|} = \sum_{l=0}^{\infty} \sum_{m=-l}^{l} \frac{4\pi }{2l+1}  \frac{r^{l}_{<}}{r^{l+1}_{>}} Y^*_{lm}(\theta^\prime, \phi^\prime) Y_{lm}(\theta, \phi),
	\label{eq:mulexp}
\end{equation}
where  $r_> (r_<)$ is the greater (smaller) of $r$ and $r^\prime$, the truncated mean field operator $W{'}^{r}_{s}  \left(\vec{r} \right)$ can be expanded as,
\begin{equation}
	W{'}^{r}_{s}  \left(\vec{r} \right) =  \sum_{l=0}^{\infty} \sum_{m=-l}^{l} V_{lm}^{rs} (r) Y_{lm}(\theta, \phi),
\end{equation}
where $V_{lm}^{rs} (r)$ is given by, 
\begin{gather}
	V_{lm}^{rs} (r) = \frac{4\pi}{2l+1} \int_{0}^{R_0} dr^\prime \frac{r^{l}_{<}}{r^{l+1}_{>}} \rho^{rs}_{lm}(r^\prime), 
	\label{eq:sR}
\end{gather}
with,
\begin{gather}
	 \rho^{rs}_{lm}(r^\prime) = \int d\Omega^\prime  Y_{lm}^{*}(\theta^\prime, \phi^\prime)  r ^{\prime 2} \psi_{r}^{*}(\vec{r^\prime}) \psi_{s}(\vec{r^\prime}).
\end{gather}
In the unscaled region ($r < R_0$), we obtain $V_{lm}^{rs} (r)$ by solving Poisson's equation \cite{McCurdy_2004},
\begin{equation}
	\left( \frac{d^2}{dr^2} - \frac{l(l+1)}{r^2} \right)( r V_{lm}^{rs} (r) )  = - 4\pi \frac{\rho_{lm}^{rs}(r)}{r}. 
\end{equation}
In the scaled region $(r > R_0)$, on the other hand, $V_{lm}^{rs} (r)$ is simplified into,
\begin{equation}
	V_{lm}^{rs} (r) = \frac{4\pi}{2l+1} \frac{1}{r^{l+1}} \int_{0}^{R_0} dr^\prime r^{\prime l} \rho^{rs}_{lm}(r^\prime),
	\label{eq:AAA}
\end{equation}
which can be evaluated by numerical integration. Hence ${W}{'} _{s}^{r} ( \vec{R}(r) )$ is expressed as,
\begin{multline}
	W{'}^{r}_{s} ( \vec{R}(r) ) =  \sum_{l=0}^{\infty} \sum_{m=-l}^{l} \frac{4\pi}{2l+1} \frac{1}{R(r)^{l+1}}  Y_{lm}(\theta, \phi) \\
	\times \int_{0}^{R_0} dr^\prime r^{\prime l} \rho^{rs}_{lm}(r^\prime).
\end{multline}

\bibliography{all8}

\end{document}